\documentclass[showpacs,superscriptaddress,twocolumn,amsmath,amssymb,prl,aps]{revtex4-1}

\usepackage{graphicx,subfigure}
\usepackage{dcolumn}
\usepackage{bm}
\usepackage{color}
\usepackage{ulem}

\newif\iffigure
\figuretrue

\newcommand{\TNS}{Ta$_2$NiSe$_5$}

\newif\ifCMbool
\CMbooltrue

\ifCMbool
\newcommand{\CMs}[1]{\textcolor{red}{{\sout{#1}}}}
\else
\newcommand{\CMs}[1]{\textcolor{red}{{ }}}
\fi

\begin{document}

\preprint{APS/123-QED}

\title{Mapping the unoccupied state dispersions in Ta$_2$NiSe$_5$ with resonant inelastic x-ray scattering}

\author{C.~Monney}
\altaffiliation{Corresponding author.\\ claude.monney@unifr.ch}
\affiliation{D{\'e}partement de Physique and Fribourg Center for Nanomaterials, Universit{\'e} de Fribourg, 1700 Fribourg, Switzerland}

\author{M.~Herzog}
\affiliation{Institut fuer Physik und Astronomie, Universitaet Potsdam, 14476 Potsdam, Germany}

\author{A.~Pulkkinen}
\affiliation{D{\'e}partement de Physique and Fribourg Center for Nanomaterials, Universit{\'e} de Fribourg, 1700 Fribourg, Switzerland}
\affiliation{School of Engineering Science, LUT University, FI-53850, Lappeenranta, Finland}

\author{Y.~Huang}
\affiliation{Swiss Light Source, Photon Science Division, Paul Scherrer Institut, 5232 Villigen PSI, Switzerland}
\affiliation{Beijing National Lab for Condensed Matter Physics, Institute of Physics, Chinese Academy of Sciences P. O. Box 603, 100190 Beijing, China}

\author{J.~Pelliciari}
\affiliation{Swiss Light Source, Photon Science Division, Paul Scherrer Institut, 5232 Villigen PSI, Switzerland}
\affiliation{NSLS-II, Brookhaven National Laboratory, Upton, NY, 11973, USA}

\author{P.~Olalde-Velasco}
\affiliation{Swiss Light Source, Photon Science Division, Paul Scherrer Institut, 5232 Villigen PSI, Switzerland}
\affiliation{Diamond Light Source, Harwell Science and Innovation Campus, Didcot, OX, OX11 0DE, UK}

\author{N.~Katayama}
\affiliation{Department of Physical Science and Engineering, Nagoya University, 464-8603 Nagoya, Japan}

\author{M.~Nohara}
\affiliation{Research Institute for Interdisciplinary Science, Okayama University, Okayama 700-8530, Japan}

\author{H.~Takagi}
\affiliation{Max Planck Institute for Solid State Research, 70569 Stuttgart, Germany}
\affiliation{Department of Physics, University of Tokyo, 113-8654 Tokyo, Japan}

\author{T.~Schmitt}
\affiliation{Swiss Light Source, Photon Science Division, Paul Scherrer Institut, 5232 Villigen PSI, Switzerland}

\author{T.~Mizokawa}
\affiliation{Department of Applied Physics, Waseda University, Shinjuku, 169-8555 Tokyo, Japan}

\begin{abstract}
The transition metal chalcogenide \TNS\ undergoes a second-order phase transition at $T_c=328 \,\text{K}$ involving a small lattice distortion. Below $T_c$, a band gap at the center of its Brillouin zone increases up to about 0.35 eV. In this work, we study the electronic structure of \TNS\ in its low-temperature semiconducting phase, using resonant inelastic x-ray scattering (RIXS) at the Ni $L_3$-edge. In addition to a weak fluorescence response, we observe a collection of intense Raman-like peaks that we attribute to electron-hole excitations. Using density functional theory calculations of its electronic band structure, we identify the main Raman-like peaks as interband transitions between valence and conduction bands. By performing angle-dependent RIXS measurements, we uncover the dispersion of these electron-hole excitations that allows us to extract the low-energy boundary of the electron-hole continuum. From the dispersion of the valence band measured by angle-resolved photoemission spectroscopy, we derive the effective mass of the lowest unoccupied conduction band.
\end{abstract}
\date{\today}

\maketitle

\section{Introduction}

The interplay of electronic correlations and structural instabilities can lead to complex phase transitions involving both the electronic and lattice degrees of freedom. In that context, the low-dimensional compound \TNS\ has stimulated large interest in the last decade, because of its tiny orthorhombic-to-monoclinic structural phase transition occurring at the critical temperature $T_c=328 \,\text{K}$ and a concomitant increase of its band gap at lower temperatures \cite{DiSalvo}. These observations launched a vivid debate on the nature and origin of this phase transition, and on the role of electron-hole interaction, in view of the possible realization of an excitonic insulator phase \cite{WakisakaPRL,Ejima}.
Thus, determining the nature of the low-energy electronic structure of \TNS\ in the normal phase, above $T_c$, and in the low-temperature phase with a reconstructed lattice, below $T_c$, is crucial to model its ground state. Indeed, the question of the existence of a Fermi surface above $T_c$ might be a key aspect to discriminate the mechanism responsible for its instability, as well as its reconstruction and possible disappearance below $T_c$. It is therefore important to gather all possible information on the electronic structure of \TNS, not only on the occupied states, but also on the unoccupied states close to the Fermi level $E_F$. Momentum-resolved probes of unoccupied states are rare and less straightforward as angle-resolved photoemission spectroscopy (ARPES) for occupied states. Recently, resonant inelastic x-ray scattering (RIXS) has been used for this purpose on the semimetal charge density wave (CDW) material TiSe$_2$ \cite{MonneyTiSeRIXS}.

\TNS\ belongs to the family of quasi-two-dimensional chalcogenides with large van der Waals gaps (see Fig. \ref{fig_1}(a)). It displays an intralayer quasi-one dimensional crystal structure with parallel chains of Ta and Ni atoms \cite{Sunshine}. Below $T_c=328 \,\text{K}$, the Ta-Ni separation shortens due to a second-order crystallographic phase transition without any signature of a CDW, and the unit cell changes from orthorhombic to monoclinic symmetry \cite{DiSalvo}. At $T_c$, the valence band maximum measured by ARPES at $\Gamma$ continuously shifts to higher binding energy up to about 0.18 eV \cite{WakisakaJSC,Seki}, and the optical band gap increases in a similar way up to about 0.16 to 0.22 eV \cite{Larkin,SeoOptics}. The nature of the low-energy electronic structure above $T_c$ is currently heavily debated, and argued to be a semimetal \cite{King,FukutaniKdoped,LeeSTM}, a zero-gap semiconductor \cite{Lu} or a semiconductor \cite{MorARPES,WakisakaJSC,Seki}. 
The difficulty of classifying its electronic structure arises from the small size of the band gap at the Fermi level and the occurrence of fluctuations of the low-temperature phase \cite{Seki}, potentially hiding the true semimetallic nature of \TNS\ \cite{Sugimoto}. In a configuration interaction picture, \TNS\ has been shown to be a negative change transfer material and its Ni site has mainly a $d^9\underline{L}$ ground state with some $d^{10}\underline{L}^2$ contribution \cite{WakisakaPRL}. In parallel, studies based on density functional theory (DFT) calculations disagree on its capability to capture the normal state of \TNS, notably due to the difficulty of finding an appropriate functional for describing this material \cite{KanekoPRBBS,LeeSTM,King,Mazza}. In addition, time-resolved studies based on pump-probe techniques support the existence of an excitonic insulator ground state at low temperature \cite{WerdehausenSciAdv, WerdehausenJPhys, OkazakiHHG,MorARPES, MorOptics}. It has been notably shown that this correlated ground state makes it possible to control the size of the band gap in \TNS, depending on the excitation density \cite{MorARPES}.
However, so far, no momentum-resolved data on the unoccupied states of \TNS\ have been published.

In this article, we present a RIXS study at the Ni $L_3-$edge of \TNS\ at low temperature, in the monoclinic phase. RIXS spectra are measured as a function of incident photon energy and they show both fluorescence and Raman-like spectral components. We focus on the Raman-like components, which we interpret as interband electron-hole excitations in the semiconducting electronic configuration of this material at low temperature. A comparison with DFT calculations permits to identify most of the low-energy RIXS peaks with specific interband transitions. By varying the in-plane component of the transferred momentum of light in the RIXS process, we map its electron-hole continuum both in energy and momentum. This allows us to extract the low-temperature band gap value of \TNS, as well as the effective mass of the lowest conduction band, thanks to a comparison with existing ARPES data.
\iffigure
\begin{figure}
\begin{center}
\includegraphics[width=8.7cm]{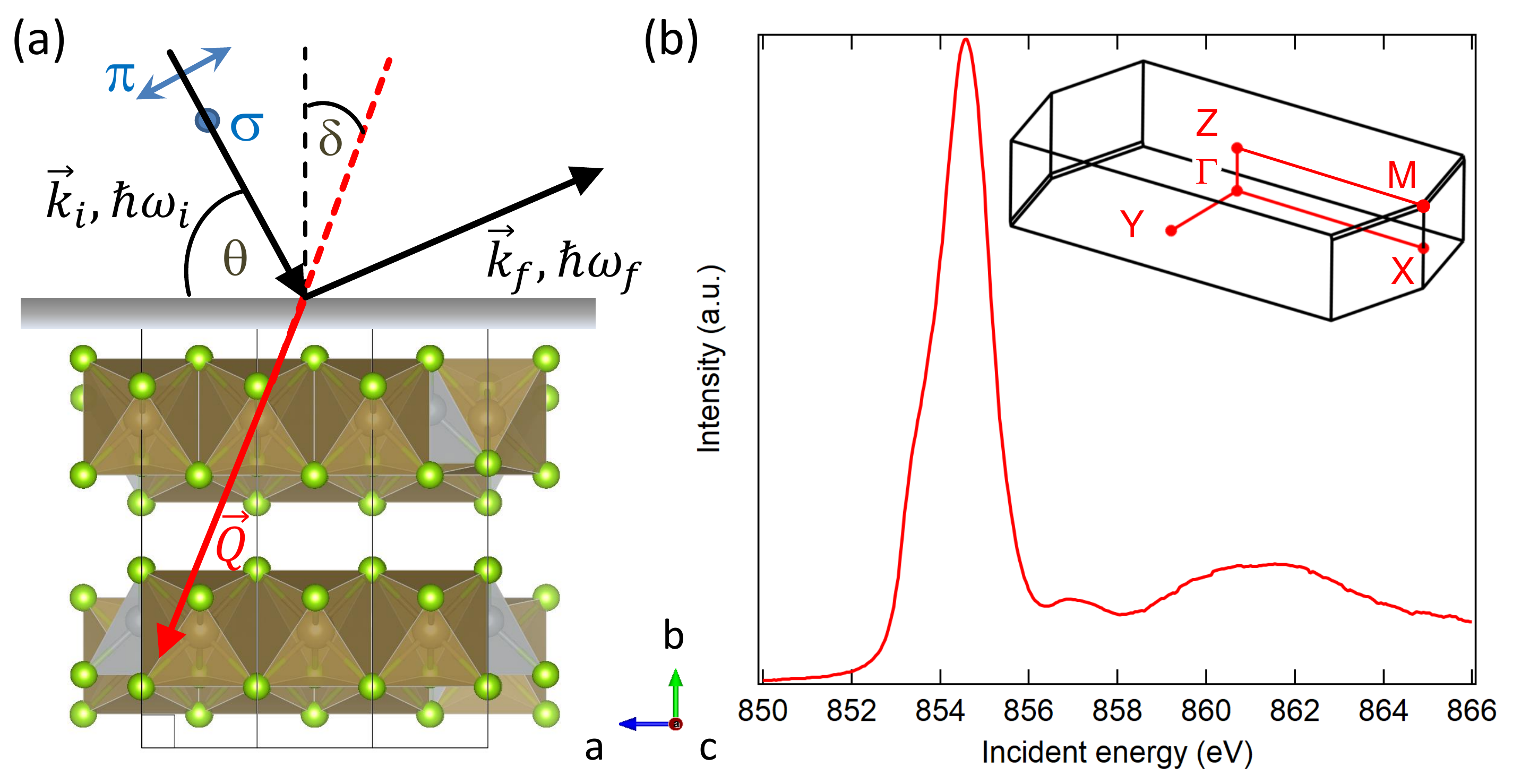}
\end{center}
\caption{\label{fig_1}
(a) Schematic description of the experimental geometry. (b) XAS spectrum of \TNS, measured at the Ni $L_3$-edge by total fluorescence yield with $\sigma-$polarized incident light at 30 K and for specular geometry ($Q_{\parallel}=0$). The Brillouin zone of the low-temperature monoclinic phase of \TNS\ is shown in the inset.
}
\end{figure}
\fi

\section{Methods}

RIXS experiments were performed at the ADRESS beamline \cite{beamline}, of the Swiss Light Source, Paul Scherrer Institut, using the SAXES spectrometer \cite{SAXES}. A scattering angle of $130^\circ$ was used and all the spectra presented here were measured with linearly $\sigma-$polarized incident light (perpendicular to the scattering plane, see Fig. \ref{fig_1} (a)). 
The combined energy resolution was 150 meV at the Ni $L_3$-edge ($\sim850$ eV). \TNS\ single crystals were cleaved in-situ (with a surface in the (001) direction) at a pressure of about $1\cdot10^{-8}$ mbar. 
\TNS\ single crystalline samples were prepared by reacting the elemental nickel, tantalum and selenium with a small amount of iodine in a evacuated quartz tube. The tube was slowly heated and kept with a temperature gradient from 950~$^{\circ}$C to 850~$^{\circ}$C for 7 days, followed by slow cooling. Single crystalline samples with a typical size of 0.04$\times$1$\times$10 mm$^3$ were obtained in the cooler end.
Density functional theory and the projector augmented wave~(PAW) method implemented in VASP~\cite{kresse1993,kresse1994,kresse1996a,kresse1996b,blochl1994,kresse1999} were used for band structure calculations in the monoclinic phase with experimentally determined structural parameters~\cite{nakano2018}. The kinetic energy cutoff of the plane-wave basis was 500~eV, and the Brillouin zone integration was performed using a $11 \times 11 \times 5$ $k$-point grid. The exchange-correlation effects were included within the strongly constrained and appropriately normed (SCAN)~\cite{sun2015} meta-generalized gradient approximation.

\section{Results and discussion}

In Fig. \ref{fig_1} (a), the geometry adopted for this study is depicted. The scattering plane (defined by the incoming and outgoing x-ray beams) is parallel to the $a-b$ planes of \TNS\ and perpendicular to the $c-$axis of \TNS. In this geometry, the transferred momentum of light $\vec{Q}$ has a component both along the $a-$axis (parallel to the direction of Ni- and Ta-chains and to the $\Gamma X$ direction in reciprocal space), $Q_{\parallel}$, and along the $b-$axis, $Q_\perp$. 
The in-plane $\Gamma Z$ and $\Gamma X$ distances are about $0.2 \,\text{\AA}^{-1}$ and $0.9 \,\text{\AA}^{-1}$, respectively, and the out-of-plane $\Gamma Y$ distance is about $0.24 \,\text{\AA}^{-1}$.
We first discuss the X-ray absorption spectroscopy (XAS) data of \TNS\ measured at the Ni $L_3$-edge in total fluorescence yield at 30 K, reported in Fig. \ref{fig_1}(b). The main structure lies in the range of energies 852 to 856 eV and consists of two peaks, the most intense one being at $854.5$ eV and the second one making a shoulder at about $853.7$ eV.
\iffigure
\begin{figure}
\begin{center}
\includegraphics[width=8.5cm]{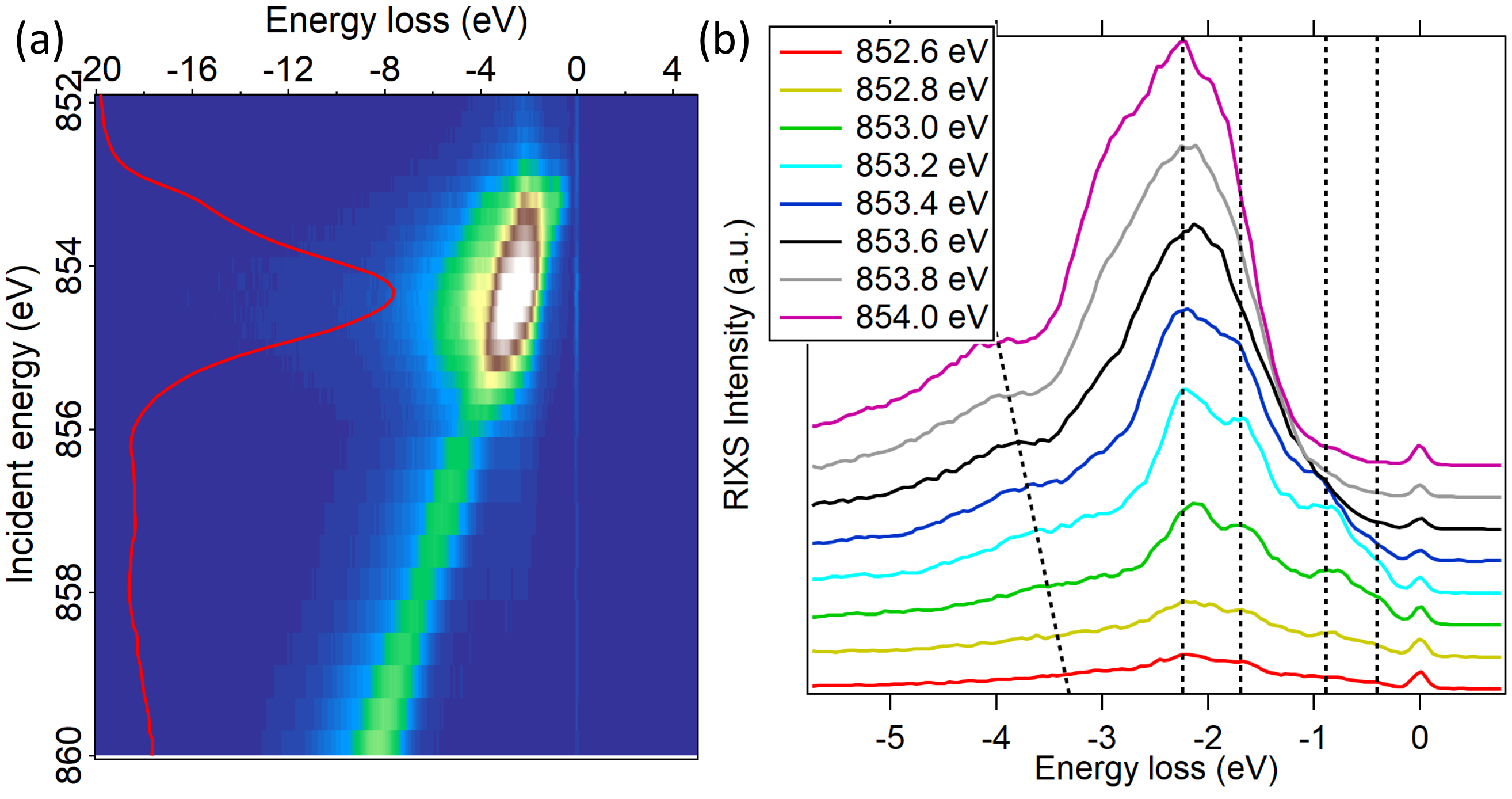}
\end{center}
\caption{\label{fig_2}
(a) RIXS map as a function of incident photon energy, measured at the Ni $L_3$-edge with $\sigma-$polarized incident light at 30 K and near specular geometry ($Q_\parallel=0.06 \,\text{\AA}^{-1}$). (b) RIXS spectra at selected incident photon energies, shown in the legend. The dashed lines indicate the main fluorescence and Raman-like structures.
}
\end{figure}
\fi

In Fig. \ref{fig_2}, we show our RIXS data obtained at $30\,\text{K}$ with $\sigma-$polarized incoming light, for incident energies $\hbar\omega_{in}$ varying along the XAS spectrum, at the Ni $L_3$-edge. The RIXS map in Fig. \ref{fig_2}(a) consists mainly of two structures: an intense broad peak centered at about $\hbar\omega_{in}=854.4 \,\text{eV}$ and a weaker peak which disperses like fluorescence from $\hbar\omega_{in}=853 \,\text{eV}$ eV up to $\hbar\omega_{in}=860 \,\text{eV}$ eV.
In Fig. \ref{fig_2}(b), we show particular RIXS spectra acquired in the low incident energy region, up to $\hbar\omega_{in}=854 \,\text{eV}$. In this range of incident photon energies, several peaks at low energy loss display a Raman behavior, i.e. appear at fixed energy loss for varied incident energies. In contrast to this, the broad fluorescence structure disperses from 3 to 4 eV energy loss. We now focus on the low energy loss Raman-like peaks and their interpretation. At low temperature, in the monoclinic phase, \TNS\ exhibits a semiconductor electronic structure with a direct gap, displaying bands dispersing on about 1 eV within $\pm 0.5$ \AA$^{-1}$ around $\Gamma$ \cite{WakisakaJSC,King}.
For these reasons, we interpret the broad peaks between 0 and 2 eV energy loss in \TNS\ as interband electron-hole excitations conserving the transferred momentum and energy of light in the scattering event of the RIXS process, in the same way as it was demonstrated for the semimetal TiSe$_2$ \cite{MonneyTiSeRIXS}. In particular, we observe that the spectra acquired for lower incident photon energies display more spectral weight at low energy loss and therefore seem to be more sensitive to low-energy electron-hole excitations.
\iffigure
\begin{figure}
\begin{center}
\includegraphics[width=8.5cm]{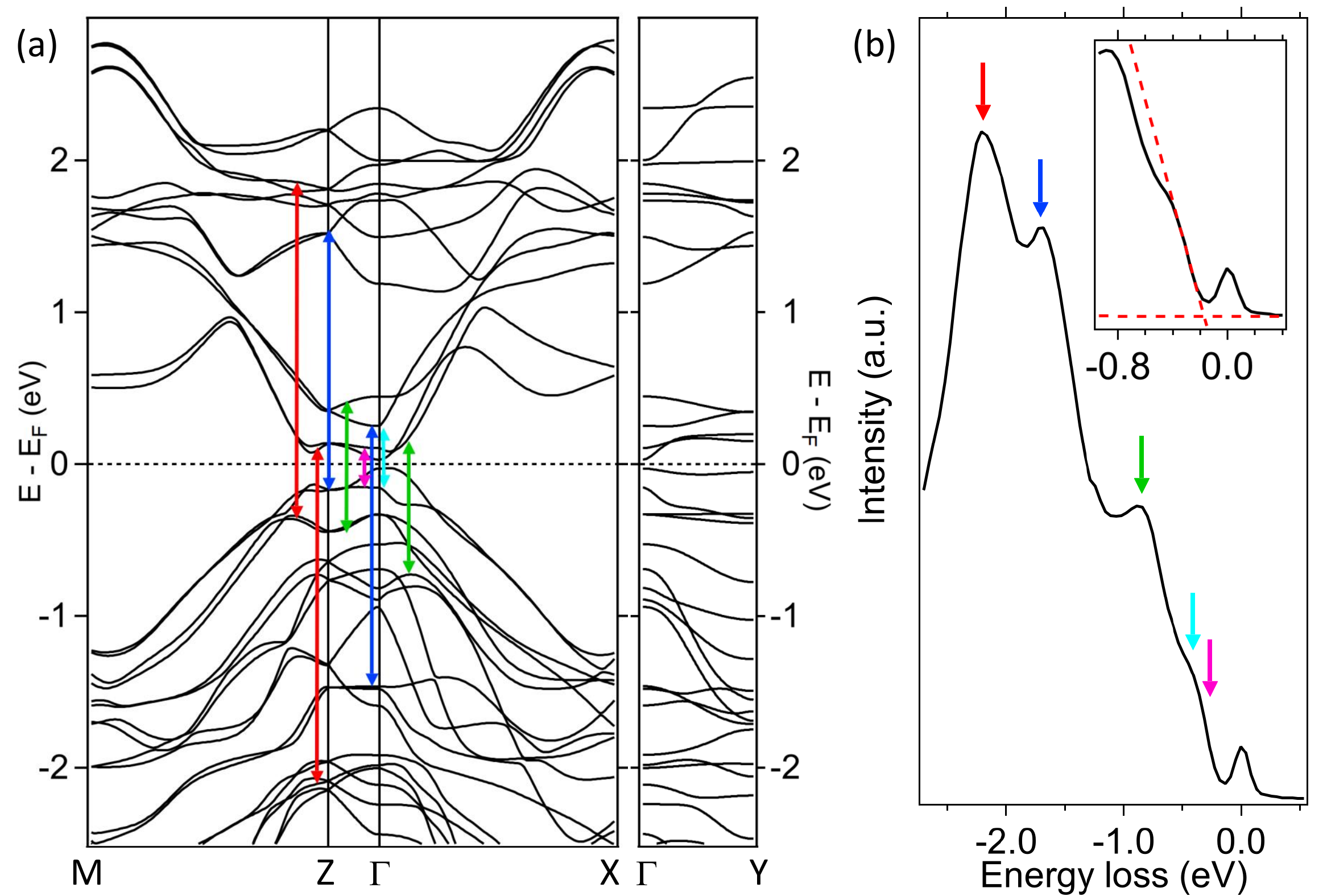}
\end{center}
\caption{\label{fig_3}
(a) Calculated band structure of \TNS\ along the main high symmetry directions (see the inset of Fig. \ref{fig_1}(b) for a description of the high symmetry points). (b) RIXS spectrum near specular geometry ($Q_\parallel=0.06 \,\text{\AA}^{-1}$) for $\hbar\omega_{in}=853.3 \,\text{eV}$ measured with $\sigma-$polarized incident light at 30 K. The colored arrows indicate fine structures in the RIXS spectrum that we attribute to specific interband transitions in the band structure of graph (a). The inset shows a possible extraction of the minimum gap by extrapolating the low-energy cut-off of the RIXS spectrum down to the baseline.
}
\end{figure}
\fi

To better interpret this, we look at the band structure of \TNS\ calculated with DFT in the low-temperature monoclinic phase, displayed in Fig. \ref{fig_3}(a). There have been various DFT studies of \TNS\ since 2012 and mainly two difficulties were encountered for this complicated material. First, the description of its normal phase band structure above $T_c$ is delicate, because the electronic structure of \TNS\ is close to being either a semimetal or a semiconductor, depending on the fine details of the DFT calculation, and especially on the choice of an appropriate functional for describing this material \cite{KanekoPRBBS,LeeSTM,King,Mazza,Subedi}. Second, the evaluation of the band gap in the low-temperature semiconducting phase is a difficult task, because it is a combination of a hybridization band gap due to the low symmetry monoclinic phase and a correlation gap due to the excitonic insulator phase. The evaluation of the hybridization gap within DFT results in a value between 30 meV and 140 meV \cite{LeeSTM,King,Subedi}. The correlation gap in the excitonic insulator phase has been so far approached by model Hamiltonians \cite{Seki,KanekoPRBBS,Ejima,Sugimoto}.
In this context, our calculated band structure agrees well with the literature, giving a hybridization gap in the monoclinic phase of about 70 meV and band dispersions similar to recent studies \cite{LeeSTM,King,Subedi}.

Based on our band structure calculation (Fig. \ref{fig_3}(a)), we attempt to attribute specific interband transitions involving occupied and unoccupied bands to the peaks and shoulders observed in the RIXS spectrum near $Q_\parallel\sim0$ below 3 eV energy loss. We display in Fig. \ref{fig_3}(b) such a RIXS spectrum taken at $\hbar\omega_{in}=853.3 \,\text{eV}$ near specular ($Q_\parallel=0.06 \,\text{\AA}^{-1}$). At this low incident energy (with respect to the main peaks in the XAS spectrum), the structures in the RIXS spectrum are the clearest and we also expect to probe electron-hole excitations close to the extrema of the valence and conduction bands.
The comparison between the calculated band structure and the RIXS spectrum is rendered in Fig. \ref{fig_3} by the colored arrows on graphs (a) and (b). The horizontal (energy) position of the vertical arrows on graph (b) are positioned approximately at the energy loss corresponding to the interband transitions in graph (a). The agreement on this assignment is good and highlights possible vertical ($Q_\parallel\sim0$) transitions that are expected to involve a large number of states in the band structure, because of relatively flat bands. We stress here that we show in Fig. \ref{fig_3} (a) only few possibilities and that this should be generalized to the whole Brillouin zone of \TNS, since all interband transitions fulfilling momentum and energy conservation are expected to contribute to the RIXS signal \footnote{in practice, integrals over the momenta of both the valence and conduction band states over the whole Brillouin zone should be done with Dirac delta functions ensuring momentum and energy conservation in the RIXS process \cite{MonneyTiSeRIXS}.}. 

In this approach, we make three assumptions. (i) We neglect the effect of band dispersions perpendicular to the surface, along $\Gamma Y$ (i.e. we neglect any influence of $Q_\perp$). This is verified in the DFT band structure plot of Fig. \ref{fig_3}(a), for which dispersions along $\Gamma Y$ are mostly below 0.2 eV (much smaller than the energy loss scale of the RIXS spectrum in Fig. \ref{fig_3}(b)). (ii) We assume that the position of the peaks and shoulders in the RIXS spectrum at lowest energy loss for $Q_\parallel=0$ is representative of direct transitions in the DFT band structure plot of Fig. \ref{fig_3}(a). Indeed, the number of possible direct transitions between two parabolic dispersions of opposite effective mass tends to increase as the energy loss (energy difference for the electron-hole excitation) increases. As a consequence, the resulting peak in RIXS becomes asymmetric on the higher energy loss side (see Fig. 3(b) in Ref. \cite{MonneyTiSeRIXS}). Again, looking at the DFT band structure plot of Fig. \ref{fig_3}(a), most dispersions near $\Gamma$ and Z are parabolic. (iii) We neglect any RIXS matrix elements modulating the intensity, as well as the orbital character of the involved states. This is an audacious assumption, but, to overcome this assumption, one would need a state-of-the-art RIXS calculation for a multiorbital dispersive material like \TNS, which goes beyond the present study.

It is instructing to note that the low-energy interband transitions resonate when the incident photon energy is tuned to the shoulder in the Ni $L_3$-edge XAS of \TNS\ at around $853.7$ eV (see Fig. \ref{fig_2}). When the photon incident energy is tuned to the main peak of the XAS, the RIXS spectra display mainly an intense peak at about -2 eV energy loss, which by analogy can be traced back to interband transitions involving the dense flat bands at around 2 eV above the Fermi level in Fig. \ref{fig_3} (a) (these states have a significant Ni character \cite{KanekoPRBBS}). This interpretation suggests that the two-peak structure of the XAS of \TNS\ is mostly reflecting the unoccupied DOS of this material, implying that the core-hole potential of the Ni $2p$ states does not have a strong effect on the XAS final states.

\iffigure
\begin{figure}
\begin{center}
\includegraphics[width=8.5cm]{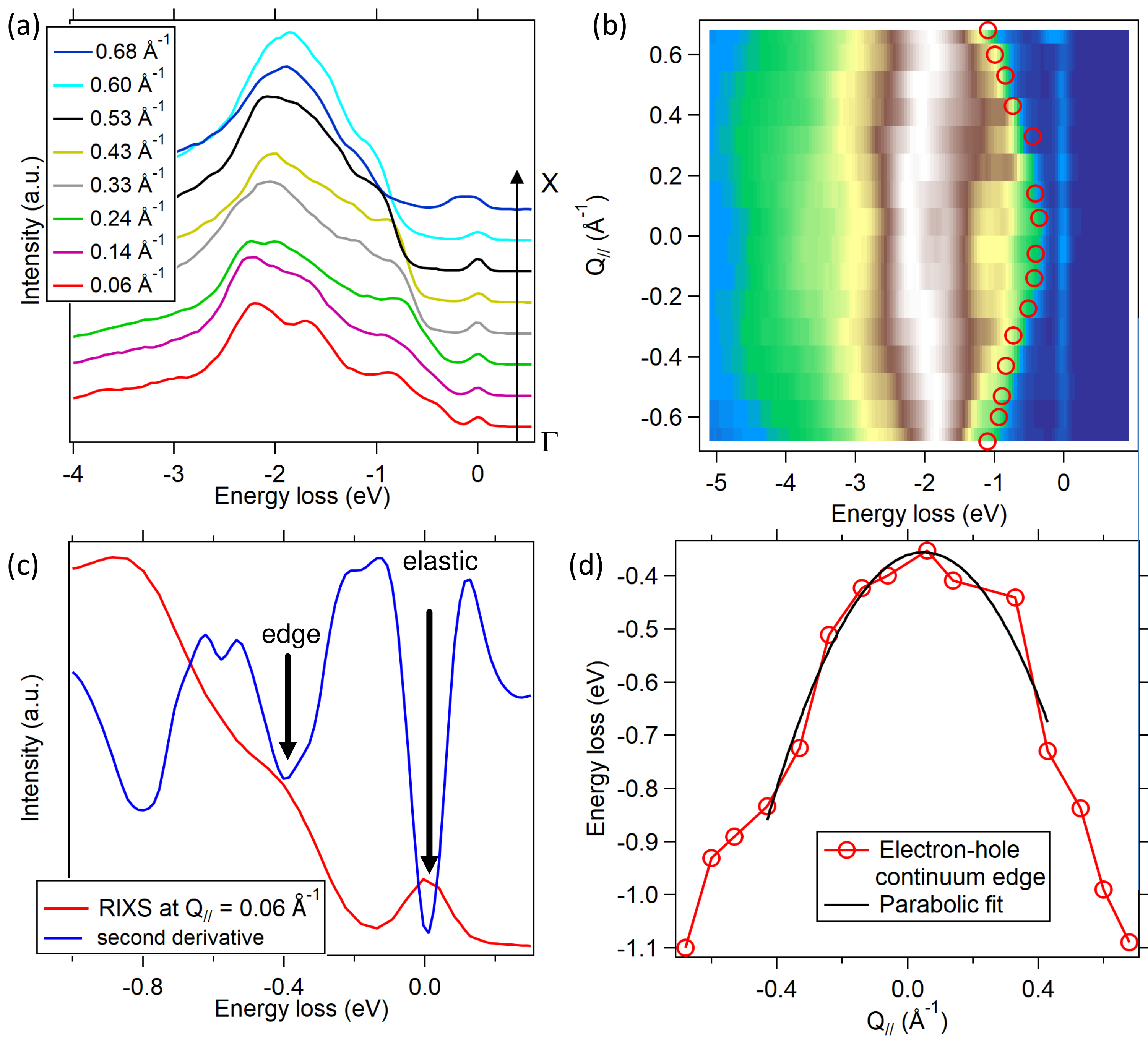}
\end{center}
\caption{\label{fig_4}
(a) RIXS spectra measured as a function of in-plane momentum $Q_\parallel$ along the $\Gamma X$ direction. These spectra were obtained for $\hbar\omega_{in}=853.3 \,\text{eV}$ and with $\sigma-$polarized incident light at 30 K. (b) RIXS map as a function of in-plane momentum $Q_\parallel$ along the $\Gamma X$ direction (same parameters as in (a)). 
(c) Example ($Q_\parallel=0.06 \,\text{\AA}^{-1}$) spectrum of graph (a) of our procedure for extracting the edge of the electron-hole continuum, using the second derivative of (smoothed) RIXS spectra. (d) Dispersion of the electron-hole continuum extracted from the RIXS map in graph (b), together with a parabolic fit near its minimum.
}
\end{figure}
\fi

Having motivated that RIXS effectively measures electron-hole excitations in \TNS, we next vary the in-plane momentum $Q_\parallel$ by rotating the sample with respect to the (fixed) incoming and outgoing light beams (in practice, this is done by varying the incident angle $\theta$ in Fig. \ref{fig_1}(a)). For negligible dispersions perpendicular to the sample surface, as confirmed above, this procedure permits to map electron-hole excitations in a momentum-resolved way. The RIXS spectra, taken at 30 K, are shown in Fig. \ref{fig_4}(a) and the corresponding false color map in Fig. \ref{fig_4}(b). Clear dispersions of the Raman-like peaks are observed, especially below 2 eV energy loss. However, it is difficult to extract all momentum-resolved contributions, because there are many overlapping peaks. We therefore focus on the low-energy edge of the RIXS spectra (below 1 eV energy loss). In our picture, it is related to the edge of the electron-hole continuum of \TNS\ and is therefore the result of the momentum- and energy-resolved convolution of the highest valence band and the lowest conduction band \cite{MonneyTiSeRIXS}.
To extract the dispersion of this continuum edge, we perform a second derivative (in the direction of the energy loss) of the smoothed RIXS spectra. An example is shown in Fig. \ref{fig_4}(c), for $Q_\parallel=0.06 \,\text{\AA}^{-1}$. We attribute the pronounced minimum with the lowest energy loss (apart from the elastic line) to the continuum edge. From the position of the lowest-energy peak for the RIXS spectra measured near specular ($Q_\parallel\sim0$), we find that the first electron-hole excitation starts at about 0.38 eV. This value corresponds to the size of the charge gap in \TNS\ at 30 K, as determined from our RIXS study, and is in good agreement with the charge gap estimated from scanning tunneling spectroscopy measurements (at 78 K) \cite{LeeSTM} or from a symmetric band behavior around the Fermi level in photoemission \cite{WakisakaJSC}. At first glance, it is substantially larger than the gap of 0.16 to 0.22 eV obtained from recent optical studies that probe $Q=0$ interband transitions \cite{Larkin,Lu,SeoOptics}. However, applying the same analysis method and extrapolating the low-energy cut-off of the RIXS spectrum down to the intensity baseline (see the inset of Fig. \ref{fig_3}(b)) provides us with a gap of 0.17 eV, in very good agreement with optical data.

Extending the second-derivative analysis to all RIXS spectra, we extract the dispersion of the continuum edge, which is plotted as open red circles in Fig. \ref{fig_4}(b). Fitting this dispersion with a parabola, we find that the effective mass of the continuum edge is about $M_{\text{tot}}=1.7\,m_e$, with $m_e$ the electron mass in vacuum. The edge of the electron-hole continuum is a convolution of the dispersions of the highest valence band and lowest conduction band. For approximate parabolic bands of effective masses $m_{\text{VB}}$ and $m_{\text{CB}}$ along the $\Gamma X$ direction, the effective mass of the convoluted dispersion measured in RIXS is $M_{\text{tot}}=m_{\text{VB}}+ m_{\text{CB}}$.
In their ARPES study of \TNS, Wakisaka \textit{et al.} have found that the effective mass of the occupied valence band along $\Gamma X$ is about $m_{\text{VB}}=0.4\,m_e$ at low temperature, based on a parabolic fit of its top part \cite{WakisakaJSC}. This value of $m_{\text{VB}}$ is most likely a lower bound, since the parabolic fit does not account for the flat top part of the valence band and is limited to a small energy range ($\sim0.3$ eV). Therefore, from the data of Wakisaka \textit{et al.} \cite{WakisakaJSC}, we estimate a higher bound for $m_{\text{VB}}=0.8\,m_e$ using a larger energy and momentum range for a parabolic fit, since the electron-hole continuum in consideration here is measured over an energy range of about 1 eV.
Using these extreme values, we find that the lowest conduction band must have an effective mass $m_{\text{CB}}$ between $1.3\,m_e$ and $0.9\,m_e$. 

The observation of clear dispersive electron-hole excitations in direct RIXS for low-dimensional materials is still quite exceptional, since it has been observed only in TiSe$_2$ \cite{MonneyTiSeRIXS} and \TNS\ so far. This is in contrast to cuprate and nickelate materials, which have attracted much attention in the last decade in the RIXS community. Especially on the underdoped side of their phase diagram, cuprates show prominent Raman-like crystal field excitations, called $dd-$excitations, that hardly disperse, except in the case of exotic orbiton physics \cite{SchlappaOrbiton,BisogniCCO,EllisCuprate}. 
In these strongly correlated systems, Eisebitt and Eberhardt have already stressed that the weakly-screened core-hole at the $L-$edge gives rise to strong excitonic effects and opens channels for relaxation of the excited electron in the intermediate state, thus making the interpretation of RIXS spectra in terms of band mapping difficult or even impossible \cite{Eberhardt}.
For TiSe$_2$ and \TNS, the screening of the core-hole is larger. In this case, the dispersive electron-hole excitations can be seen as dispersive $dd-$excitations (or crystal field excitations), in the sense that they are the result of energy- and momentum-conserving transitions between electronic bands split by the crystal field (and by their dispersion due to transfer integrals).
However, electron-hole excitations appear on the overdoped side of cuprates and their modeling with the charge susceptibility \cite{MonneyLSCO,SuzukiNPJ,Guarise} (or more evolved theoretical approaches \cite{Demler1,Demler2}) tends to attest of a similar picture as the one described in this work.

In that framework, it is remarkable that such dispersive electron-hole excitations are not observed in iron pnictides \cite{ZhouPnict,PelliPnict,PelliComm,NomuraPnict,PelliPRB,Rahn,Hancock,PelliAPL}. The main difference in that respect is that pnictides are semimetals with a rather large density of states near the Fermi level, contrarily to \TNS\ and TiSe$_2$ (which is a semimetal with a very small density of states near the Fermi level). Additionally, magnetic excitations are dominant in RIXS for pnictides at low energy, while both \TNS\ and TiSe$_2$ are non-magnetic materials. We believe that the presence of many charge carriers near the Fermi level makes it easy for the excited valence electron in the intermediate state of the RIXS process to scatter due to electron-electron interaction, leading to a loss of momentum conservation between the excited electron and hole in the final state of the RIXS process. This is described as $k$-unselective processes (or sometimes incoherent processes) \cite{Eberhardt}. It would be therefore very interesting to extend this momentum-resolved measurements of dispersive electron-hole excitations to semiconducting materials with large band gaps, to better determine the necessary conditions for (coherent) $k$-selective processes.

\section{Conclusions}

In summary, we have measured the quasi-one-dimensional material \TNS\ with resonant inelastic x-ray scattering at the Ni $L_3-$edge at low temperature. In addition to a fluorescence component, we observe Raman-like excitations up to 2 eV energy loss for incident photon energies tuned to a pre-edge shoulder in the X-ray absorption spectrum. They are interpreted as interband electron-hole excitations in this dispersive material and are identified as specific transitions between bands calculated with density functional theory. Focusing on these interband excitations, we have acquired RIXS spectra as a function of parallel momentum along the most dispersive axis of \TNS. In this way, the momentum- and energy-resolved electron-hole continuum of \TNS\ is mapped. This allows us to estimate the band gap of this semiconductor and also to extract the effective mass of the lowest conduction band from the edge of the dispersive electron-hole continuum. We hope that our experimental work will serve as a benchmark for a delocalized and weakly correlated material and will stimulate a state-of-the-art RIXS calculation like the new dynamical mean-field theory based method recently proposed by Hariki {\it et al.} \cite{DMFT}.

\section{Acknowledgments}

This project was supported from the Swiss National Science Foundation (SNSF) Grant No. P00P2\_170597. A.P. acknowledges the Osk. Huttunen Foundation for financial support, and CSC-IT Center for Science, Finland, for computational resources. J.P. and T.S. acknowledge financial support through the Dysenos AG by Kabelwerke Brugg AG Holding, Fachhochschule Nordwestschweiz and the Paul Scherrer Institut. Work at PSI has been funded by the Swiss National Science Foundation through the Sinergia network Mott Physics Beyond the Heisenberg (MPBH) model (SNSF Research grant numbers CRSII2\_141962 and CRSII2\_160765). The research leading to these results has received funding from the European Community’s Seventh Framework Programme (FP7/2007$-$2013) under Grant Agreement No. 290605 (COFUND: PSIFELLOW). The experiment was performed at the ADRESS beamline of the Swiss Light Source at the Paul Scherrer Institut.

\end{document}